\documentstyle[aps,twocolumn,prl,epsf]{revtex}
\begin{document}
 
\def\e{\varepsilon}
\def\ec{\varepsilon_c}
\def\ds{\displaystyle}
\def\cml{coupled map lattice}
\def\cmls{coupled map lattices}
\def\CML{Coupled Map Lattice}
\def\CMLs{Coupled Map Lattices}
\font\eusm=eusm10
\def\WSS{World Scientific Singapore}
\def\CUP{Cambridge Univ. Press}
\def\OUP{Oxford Univ. Press}
\def\SV{Springer-Verlag, Berlin}
\def\AW{Addison-Wesley}
\def\PTP{Prog.\ Theor.\ Phys.}
\def\PLA{Phys.\ Lett.\ A}
\def\PRE{Phys.\ Rev.\ E}
\def\PRA{Phys.\ Rev.\ A}
\def\PD{Physica D}
\def\JSP{J.\ Stat.\ Phys.}
\def\MPCPS{Math.\ Proc.\ Camb.\ Phil.\ Soc.}
\def\JCSFT{J.\ Chem.\ Soc.\ Faraday.\ Trans.}
\def\McGH{Mc Graw-Hill}
\def\Kan{K.\ Kaneko}

\input amssym.def
\input amssym.tex

\draft

\title{Mode-locking in Coupled Map Lattices}
\author{R.\ Carretero-Gonz\'alez\thanks{e-mail: R.Carretero@ucl.ac.uk}
        \thanks{New address: Center for Nonlinear Dynamics and its 
        Applications (CNDA), Dept.~Civil and Environmental Engineering,
        University College London, Gower Street, LONDON WC1E 6BT},
	D.K.\ Arrowsmith and F.\ Vivaldi}

\address{
	School of Mathematical Sciences, 
	Queen Mary and Westfield College,
	Mile End Road, London E1 4NS, U.K.}

\date{Physica D {\bf 103} (1997) 381--403}

\maketitle


\begin{abstract}
We study propagation of pulses along one-way {\cmls}, 
which originate from the transition between two superstable 
states of the local map. The velocity of the pulses exhibits a 
staircase-like behaviour as the coupling parameter is varied. 
For a piece-wise linear local map, we prove that the velocity of the 
wave has a Devil's staircase dependence on the coupling parameter.
A wave travelling with rational velocity is found to be stable to
parametric perturbations in a manner akin to rational mode-locking 
for circle maps. We provide evidence that mode-locking is also
present for a broader range of maps and couplings.
\end{abstract}

\section[Introduction]{Introduction}

An interesting feature of {\cmls} (CML) \cite{Kan:book,Atlee:book}
is the widespread occurrence of the so-called 
{\it spatio-temporal chaos\/} \cite{Kan:89,Keeler:86}, 
which is irregular behaviour in time as well as space. 
Equally interesting is the appearance of coherent 
structures from an apparently decorrelated medium 
\cite{Buni:90}--\nocite{Ying:94}\cite{Kapral:86}. For example, 
an interface separating two different phases may travel along 
the lattice \cite{Kapral:94}--\nocite{Kan:93,Amrit:91}\cite{Gade:93}. 
The movement of such a front depends on the strength of the 
coupling between lattice sites. This paper is concerned with 
investigating some aspects of this phenomenon.

We consider an infinite collection of cells with a local 
dynamical variable $x_t(i)$, characterizing the state of 
the $i$-th cell at time $t$. Both $i$ and $t$ are integer.
The global state of the lattice at time $t$ is given by the 
state vector 
\begin{equation}
\label{GlobalState}
X_t=\{x_t(i)\}=(\ldots,x_t(i-1),x_t(i),x_t(i+1),\ldots).
\end{equation}
The state of the $i$-th cell at time $t+1$ given by
\begin{equation} \label{CML0} 
x_{t+1}(i)=F_i(\dots,x_t(i-1),x_t(i),x_t(i+1),\dots),
\end{equation}
where the function $F_i$ is determined by the local dynamics 
in each cell as well as the interaction between cells. 
A physically meaningful interaction will have typically a 
limited range, with decreasing strength for distant neighbours. 
Equation (\ref{CML0}) defines the components of the global map $F$: 
$$
F=(\,\ldots,F_{-1},F_0,F_1,\ldots\,)\qquad X_{t+1}=F(X_t).
$$
A mapping of global states that will be of use in the
following is the {\it shift mapping\/}
\begin{equation} \label{ShiftMapping}
G\left(\{x_t(i)\}\right)\,=\,\{x_t(i+1)\}.
\end{equation}
It should be clear that $F$ and $G$ commute.

We consider {\it homogeneous\/} CML (same map $F_i$ for all cells) 
with linear nearest-neighbours interaction. The two most widely used 
models are
\begin{equation} \label{diffu}
x_{t+1}(i)= (1-\e) f(x_t(i)) + {\e\over 2}
	   \left( f(x_t(i-1)) + f(x_t(i+1)) \right)
\end{equation}
and
\begin{equation} \label{one_way}
x_{t+1}(i)= (1-\e) f(x_t(i)) + \e f(x_t(i-1)),
\end{equation}
which are called {\it diffusive\/} and {\it one-way\/} CML respectively. 
The coupling strength $\e$ (also called the {\it diffusive parameter\/})
satisfies $0\leq\e\leq 1$ to ensure the conservation of flux between
neighbouring sites.

In this paper we consider the propagation of localized wavefronts
along a linear one-way CML. Localized states and step states 
(localized wavefronts) are introduced in the next section, where 
we establish some of their basic properties. In section 3 we 
introduce a one-parameter family of piece-wise linear local maps, 
for which wavefronts can propagate coherently in a simple way. In 
the following section we show that the dynamics of an important 
class of wavefronts can be reduced to the study of a discontinuous 
piecewise-linear circle map, depending on two parameters. We study 
this family with methods of symbolic dynamics. These results are 
then applied to the pulse propagation in section 5. We show that 
for each value of the parameters the velocity of the pulse is 
well-defined, and we use it to characterize the structure of the 
parameter space. The latter is found to be structurally similar to 
that of a subcritical circle map, with velocity playing the role of 
rotation number. We establish the occurrence of a critical transition 
from a stable standing signal to a stable travelling one, as the 
coupling parameter is increased, and we prove parametric stability 
of pulses travelling with rational velocity. The velocity of the 
travelling pulse is found to have Devil's staircase-like dependence 
on the coupling parameter (see figure 1). In the last section we 
indicate how to generalize our results to a broader range of functions 
and couplings.
\bigskip

\section[2]{Localized states and propagating fronts}

We first define localized states, step states and their 
velocity of propagation and then establish some of their 
basic properties. In what follows, we deal with a linear 
one-way CML, unless stated otherwise.
\medskip

\subsection[2.1]{Localized states and step states}

Let $\Delta x_t(i)= |x_t(i+1)-x_t(i)|$. The {\it mass\/} $M(X_t)$ of 
a global state $X_t$ is defined as the variation of the local states
$$
M_t=M(X_t)= \sum_{i=-\infty}^\infty \Delta x_t(i).
$$
We restrict our attention to states with positive finite mass
(a zero-mass state is a uniform state).

For each lattice site $i\in\Bbb Z$ we define the probability
\begin{equation} \label{Probability}
 p_t(i)\,=\, 
    {\frac{\ds \Delta x_t(i)}
     {\ds M_t}}.
\end{equation}
The mean $\mu$ and the variance $\sigma$ of the variable $i$ with 
respect to the probability distribution (\ref{Probability}) are given by
$$\begin{array}{l}
\mu_t=\mu(X_t)= \ds\sum_{i=-\infty}^\infty i\,p_t(i)\\[2.0ex]
\sigma^2_t=\sigma(X_t)^2
	  =\ds\sum_{i=-\infty}^\infty (i-\mu_t)^2\,p_t(i).
\end{array}$$
The quantity $\mu_t$ is called {\it centre of mass\/} of the state 
$X_t$, and $\sigma_t$ its {\it width}. The centre of mass and
the width give, respectively, the average position and spread of the 
localized state.

A state $X_t$ with finite centre of mass and width is said to be 
{\it localized}. If, in addition, there exists a positive constant 
$\kappa <1$ such that $\Delta x_t(i)=O(\kappa^{|i|})$, then the 
state is {\it exponentially localized}. For an exponentially 
localized state, not only the mean and variance exist, but also all 
central moments of the distribution (\ref{Probability}).

We are interested in localized states that can model wavefronts.
So we define a {\it step state\/} to be a localized state for which 
the configuration $\{x_t(i)\}$ is asymptotic, for $i\to\pm\infty$, 
to two distinct fixed points of $f$ \cite{Bastien:95}:
\begin{equation} \label{FixedPoints}
\lim_{i\to\pm\infty} x_t(i)= x^*_\pm; \qquad f(x^*_\pm)=x^*_\pm.
\end{equation}
The mass of a step state is bounded from below 
\begin{equation}\label{MinimalMass}
M(X_t)\geq \Delta x^*=|x^*_+-x^*_-|.
\end{equation}
The simplest such state is a {\it pure step state},
that is, a step function between the two fixed points
\begin{equation} \label{PureStepState} 
    P(k)\equiv x_t(i)=\left\{ \begin{array}{l} 
	x_-^*\quad\mbox{if}\quad i\leq k\\[1.0ex]
	x_+^*\quad\mbox{if}\quad i>k.
\end{array}\right.\end{equation}
For this pure state $\mu(P(k))=k$ and $\sigma(P(k))=0$.

A step state has {\it minimal mass\/} or {\it minimal variation\/},
{\it i.e.}\ $M(x_t)=\Delta x^*$ 
(cf.~(\ref{MinimalMass})), if and only if its local 
states are ordered, that is if $x_t(i)\leq x_t(i+1)$ when $x_-^*<x_+^*$ 
(or $x_t(i)\geq x_t(i+1)$ when $x_-^*>x_+^*$) for all $i$. It is plain 
that if the configuration of the one-way CML has minimal mass at time $t$ 
it is going to preserve its minimal mass at any time greater than $t$ 
provided that $f$ is a non-decreasing function if $x_-^*<x_+^*$ 
and non-increasing if $x_-^*>x_+^*$.

The global dynamics causes centre of mass and width of a localized 
state to evolve. In this respect, two points need consideration. 
Firstly, the existence of the probability $p_t$ does not automatically 
imply that of $p_{t+1}$, and secondly, the image of a localized state 
is not necessarily localized. To achieve the above conditions, we impose 
some mild restrictions on the local map $f$ as well as the state $X_t$.
The importance of the states defined above is that they are invariant 
under the dynamics of the CML --- for the complete statements and proofs 
see appendix A. This gives the framework for the existence of steady 
step states and travelling ones.
\medskip

\subsection[2.2]{Travelling velocity and basic properties}

From the previous definitions and results it is clear that if the initial 
state $X_0$ is a step state, then $\mu(X_t)$ and $\sigma(X_t)$ are defined 
for all $t\geq 0$ (see appendix A for details). Our main interest is to 
determine the average velocity $v$ of the centre of mass along a global 
orbit, namely 
$$
v\,=\,v(\e, X_0)\,=\,\lim_{t\to\infty} {1\over t} (\mu_t - \mu_0).
$$
A formal analysis on the stability of exponentially localized
step states establishes the following quite general property of
the velocity of a step state in one-way CML:
\medskip

\proclaim Theorem 1. Let $X_0$ be an exponentially localized step 
state of a one-way CML. Let the local map $f$ be bounded, and let 
$f$ be a contraction mapping in a neighbourhood of the fixed points 
{\rm (\ref{FixedPoints})}. Then, for all sufficiently small $\e$,
$v(\e)=0$ and $v(1-\e)=1$, independently of $X_0$.
\medskip

The proof of this theorem is given in appendix A.
The condition that $f$ be a contraction mapping is a bit 
stronger than the condition that $x^*_\pm$ be attracting.
We shall see that for an initial localized 
interface of the type (\ref{PureStepState}), the critical 
behaviour is much richer than a discontinuous transition 
from $v=0$ to $v=1$, at some intermediate value of $\e$.
Rather, the dependence of $v$ on $\e$ is characterized
by infinitely many critical values, in correspondence to 
the boundaries of intervals in $\e$, where the velocity is 
a given rational number (see figure 1).

A general feature of the velocity of the travelling wave is 
its symmetry. Let us define $\delta = 1-\e$ and consider a 
moving reference frame with a unit positive velocity:
$$ \left\{ \begin{array}{ccl}
   y_t(i)&=&x_t(i) \\[1.0ex]
   y_{t+1}(i)&=&x_{t+1}(i+1) \\
   \vdots\quad& & \quad \vdots\\
   y_{t+k}(i)&=&x_{t+k}(i+k).
\end{array} \right. $$
The one-way CML in the moving frame now reads
\begin{equation} \label{one_way'} \begin{array}{rccl}
	     &y_{t+1}(i-1)&=&\delta f(y_t(i))+(1-\delta)f(y_t(i-1))\\[1.5ex]
\Longrightarrow&y_{t+1}(i)&=& (1-\delta)f(y_t(i)) + \delta f(y_t(i+1)).
\end{array} \end{equation}
Equation (\ref{one_way'}) represents a one-way CML of the type
(\ref{one_way}) but coupled in the opposite direction. Thus, taking into
account the moving reference frame and the definition of $\delta$ we can 
conclude that 
\begin{equation} \label{Symmetry}
1-v(\e)=v(1-\e).
\end{equation} 
Therefore the velocity of the travelling wave in a one-way CML of the form 
(\ref{one_way}) is symmetric with respect to the point $(\e,v(\e))=(1/2,1/2)$ 
(see figure 1). This important symmetric property allows one to restrict the
study of the velocity to the $\e$-interval $[0,1/2]$.
\bigskip

\section[3]{A piece-wise linear local map}

In this section we deal with the coherent propagation of a 
wavefront (without damping or dispersion) along the lattice. 
We confine our attention to states with minimal mass. In order 
to ensure that this property is preserved under iteration, we 
shall require the local map $f$ to be non-decreasing as stated 
before. In addition, we ask $f$ to be continuous and to have two 
stable fixed points $x_-^*$ and $x_+^*$ (with $x_-^*<x_+^*$). 
So $f$ will have precisely one unstable fixed point at $x^*$ 
with $x_-^*<x^*<x_+^*$. The local dynamics takes place on 
the interval $I=[x_-^*,x_+^*]$, where there are two basins of 
attraction $I_-=[x_-^*,x^*)$ and $I_+=(x^*,x_+^*]$ for the fixed 
points $x_-^*$ and $x_+^*$, respectively.
\medskip

\subsection[3.1]{The local mapping}

A simple one-parameter family of maps that fulfills the above 
requirement is given by (see figure 2)
\begin{equation} \label{fa} f_a(x)=\left\{ \begin{array}{cll} 
				       -1&\mbox{if}& x\leq -a\\[1.2ex]
{\ds 1\over \ds a}\,x&\mbox{if}& -a<x<a\\[1.2ex]
				       1 &\mbox{if}& x\geq a
\end{array}\right.\qquad 0 < a < 1,\end{equation}
and by the step function between $-1$ and $1$ centered at 
the origin when $a=0$. The fixed points $x_-^*=-1$ and $x_+^*=1$ are 
superstable, with basins $I_-=[-1,0)$ and $I_+=(0,1]$, respectively. 
The repeller is the origin $x^*=0$. The coupled map lattice now depends 
on the two parameters $\e$ and $a$, and the parameter space is the unit 
square. The one-way CML with local mapping $f_a$ and coupling $\e$ will 
be denoted by $F_{\e,a}$.

We partition the interval $I=[-1,1]$ into the unstable domain $U=(-a,a)$ 
and the superstable domains $S_-=[-1,-a]$ and $S_+=[a,1]$. Any site 
falling within $S_\pm$ maps, under $f_a$, to $x^*_\pm$ at the next
iteration. 

We begin considering states of minimal mass $X_t$ whose configurations
$\{x_t(i)\}$ form a {\it non-decreasing\/} sequence. Because $X_t$ is a 
localized state, only a finite number $n$ of local states belong to $U$:
\begin{equation}\label{Range}
i\in \{k+1,\ldots,k+n\} \quad\Longleftrightarrow\quad x_t(i)\in U.
\end{equation}
For $i\leq k$ ($i>k+n$), the local states belong to the 
superstable domain $S_-$ ($S_+$) and their image under 
$f_a$ is $-1$ ($+1$). Consequently, if we let $X_t'$ be 
the state obtained from $X_t$ by replacing $x_t(i)$ with 
$-1$ for $i\leq k$ and with $+1$ for $i>k+n$, then 
$F_{\e,a}(X_t)=F_{\e,a}(X_t')$. The state $X_t'$ is the 
{\it reduced state\/} associated with $X_t$, and we write 
$X_t\sim X_t'$. We will reduce all states after each 
iteration of the mapping $F_{\e,a}$.

So, without loss of generality, we shall restrict our attention 
to states of the type
$$ \begin{array}{l}
X_t=(\,\ldots, -1,-1,x_t(k+1),\ldots,x_t(k+n),1,1,\ldots\,)\\[1.0ex]
\qquad{\rm with} \quad \bigl|x_t(i)\bigr|<a,\quad i=k+1,\ldots,k+n.
\end{array}$$
The number $n$ of local states in $U$ is called the {\it size\/} 
of $X_t$. The knowledge of the range (\ref{Range}) provides useful 
information about position and width of a step state. Indeed the 
center of mass of $X_t$ satisfies the bounds
\begin{equation}\label{EstimateForMu}
k+{n(1-a)\over 2} < \mu(X_t) < (k+n)-{n(1-a)\over 2}
\end{equation}
and the width
\begin{equation}\label{EstimateForSigma}
{n(1-a)\over 2}<\sigma(X_t) <{n\over 2}.
\end{equation}

The image of $X_t$ is
$$
\begin{array}{rcl}
X_{t+1}&=&(\ldots,-1, -1,f_-(x_t(k+1)),\ldots,\\
       & &x_{t+1}(k+n),f_+(x_t(k+n)),1,\ldots\,)
\end{array}
$$
where
\begin{equation} \label{fPlusMinus} \begin{array}{cl}
   f_-(x)=&{\ds 1-\e\over \ds a}\,x-\e\\[1.8ex]
   f_+(x)=&{\ds \e\over \ds a}\,x + (1-\e).
\end{array} \end{equation}
$X_{t+1}$ has at most $n+1$ sites in $U$. The reduced state $X'_{t+1}$ 
is one of the following:
\setlength{\arraycolsep}{1pt} 
\begin{equation}\label{NextState} \begin{array}{llcccl}
(a) & (\ldots,-1, -1,&x_{t+1}(k+1),  &x_{t+1}(k+2),\ldots,\\
    & x_{t+1}(k+n),  &1,             &1,\ldots\,) \\[1.2ex]
(b) & (\ldots,-1, -1,&-1,            &x_{t+1}(k+2),\ldots,\\
    & x_{t+1}(k+n),  &x_{t+1}(k+n+1),&1,\ldots\,) \\[1.2ex]
(c) & (\ldots,-1, -1,&x_{t+1}(k+1),  &x_{t+1}(k+2),\ldots,\\
    & x_{t+1}(k+n),  &x_{t+1}(k+n+1),&1,\ldots\,)
\end{array} \end{equation}
Loosely speaking, as the time increases from $t$ to $t+1$, the state 
has not propagated in case (a), it has advanced to the right by one 
site in case $(b)$ and it has propagated with dispersion in case (c).

To decide among these possibilities, one needs to investigate 
the value of $f_-$ and $f_+$ at the boundary sites $i=k+1$ and 
$i=k+n$, respectively. Letting
$$
\gamma_-={a(\e - a)\over 1-\e}\qquad \gamma_+={a(a+\e-1)\over\e}
$$
one verifies that the functions $f_\pm$ satisfies the following equations
\begin{equation} \begin{array}{ll} \label{LotsOfValues}
 f_-(-a) = -1     &\quad f_+(-a) = 1-2\e \\[1.0ex]
 f_-(a) = 1-2\e   &\quad f_+(a) = 1 \\[1.0ex]
 f_-(\gamma_-)=-a &\quad f_+(\gamma_+)=a.
\end{array} \end{equation}

We define the {\it gap\/} of the CML with local map $f_a$ to be the
open interval $\Gamma=(\gamma_+,\gamma_-)$. Its length (with sign) 
is given by
\begin{equation}\label{GapSize}
\gamma(\e,a)\,=\,\gamma_--\gamma_+
\,=\,{\frac{\ds a(1-a-2\e(1-\e))}{\ds\e(1-\e)}}.
\end{equation}

Let $x\in U$. 
Then $f_+(x)\in U$ if $x \leq \gamma_+$, and $f_-(x)\in U$
if $x\geq\gamma_-$. From this and the fact that $x_t(i)$ is 
a non-decreasing function of $i$ we obtain that the size of 
a step state of minimal mass cannot increase, unless it is 
zero, for a gap of non-negative size.

In particular, it is not difficult to prove that when the gap 
size is non-negative, if the size is not greater than 1 at time 0, 
then it will remain so for all times. A {\it minimal state\/} is a 
minimal mass step state of size 0 or 1. Therefore, if we start with 
a minimal state, the configuration of the CML will remain a minimal 
state at any time.
\bigskip

\section[4]{Symbolic dynamics of minimal states}

We develop a symbolic description of the dynamics of minimal states in the 
case of non-negative gap length. Their dynamics can be reduced to the iteration
of a one-dimensional piecewise linear circle map --- the auxiliary map.
\medskip

\subsection[4.1]{The auxiliary map}

Without loss of generality, we consider minimal states of the form
\begin{equation} \label{MinimalState} 
X_t = (\dots,-1,-1,x_t(i),1,1,\dots),
\end{equation}
where either $|x_t(i)|\leq a$ or $x_t(i)=-1$,
for which we introduce the shorthand notation
$$
X_t=[x,i\/]_t,
$$
(the subscript $t$ will often be dropped).
Thus $[-1,i\/]=P(i)$ --- cf.~(\ref{PureStepState}).
The image of $X_t$ is given by
\begin{equation} \label{X(t+1)}
X_{t+1} = (\dots,-1,-1,f_-(x_t(i)),f_+(x_t(i)),1,\dots).
\end{equation}

We define
\begin{equation} \label{EpsilonCritical}
\ec = {\ds 1-a\over \ds 2}
\end{equation}
and note that $\ec$ is the area enclosed between the graph of 
$f_a$ and the diagonal, for $x^*<x<x_+^*$. The following two results 
characterize completely the evolution of $X_t$ in the parameter 
ranges $\e \leq \ec$ and $\e \geq 1- \ec$.
\medskip

\proclaim Theorem 2. Let $X_t$ be a minimal state of
the form {\rm (\ref{MinimalState})}, and let
\begin{equation} \label{x_pm}
x_-={\e a\over 1-\e -a}\qquad x_+={a(1-\e) \over a-\e}.
\end{equation}
(i)\quad  If $\e \leq \ec$, then for $x(i)<x_-$ ($x(i)>x_-$) the 
state $[x,i\/]$ reaches $P(i)$ ($P(i-1)$) in a finite time. The 
state $[x_-,i\/]$ is fixed and unstable.
\vskip5pt
\noindent (ii)\quad  If $\e \geq 1-\ec$, then for $x(i)<x_+$ 
($x(i)>x_+$) the state $[x,i\/]$ reaches $P(i+k)$ ($P(i+k-1)$) in 
a finite time. The state $[x_+,i\/]$ is fixed and unstable under 
$G\circ F$, where $G$ is the shift mapping {\rm(\ref{ShiftMapping})}.
\medskip

\noindent The proof of the above results is given in appendix B. From 
Theorem 2 it follows that the velocity of the travelling interface satisfies
\begin{equation}\label{ZerothPlateaus} v(\e)=\left\{\begin{array}{ll}
0 &\quad{\rm if}\quad 0\leq\e\leq\ec\\[1.0ex]
1 &\quad{\rm if}\quad 1-\ec\leq\e\leq 1.
\end{array}\right. \end{equation}
As a consequence, we have that if $\e \leq \ec$ then $F(P(i))\sim P(i)$,
and if $\e\geq 1-\ec$ then $F(P(i))\sim P(i+1)$. In the rest of this 
chapter we shall assume that $\ec < \e <1-\ec$.

We partition $U$ into three intervals, namely
$$ \begin{array}{ll}
U_+&=\,\{x: -a\leq x \leq \gamma_+\}\\[1.0ex]
\Gamma &=\,\{x: \gamma_+< x < \gamma_-\}\\[1.0ex]
U_-&=\,\{x: \gamma_-\leq x \leq a\}.
\end{array}$$
Because we are assuming the gap length to be non-negative, we are 
dealing with a minimal state and then the three possibilities for
the state (\ref{X(t+1)}) are:
\begin{equation} \label{ThreePossibilities}
\setlength{\arraycolsep}{2pt}\begin{array}{lccl} 
 (a)& x_t\in U_-  &\quad\Rightarrow&\quad f_-(x_t)\in U,  \,f_+(x_t)\in S_+
\\[1.0ex]
(b')& x_t\in\Gamma&\quad\Rightarrow&\quad f_-(x_t)\in S_-,\,f_+(x_t)\in S_+
\\[1.0ex]
 (b)& x_t\in U_+  &\quad\Rightarrow&\quad f_-(x_t)\in S_-,\,f_+(x_t)\in U.
\end{array} \end{equation}
corresponding to the reduced states --- see (\ref{NextState})
$$\setlength{\arraycolsep}{1pt} \begin{array}{llccl}
 (a)&\quad X_{t+1} =(\ldots,-1,-1,&x_{t+1},&1,&1,\ldots)\\[1.2ex]
(b')&\quad X_{t+1} =(\ldots,-1,-1,&-1     ,&1,&1,\ldots)\\[1.2ex]
 (b)&\quad X_{t+1} =(\ldots,-1,-1,&-1     ,&x_{t+1},&1,\ldots)
\end{array}$$
For the case $(b')$ we have the evolution
\begin{equation} \label{TwoStages}
[x,i]_{t}\,\longrightarrow \,[-1,i]_{t+1}\,
	       \longrightarrow \,[1-2\e,i+1]_{t+2}.
\end{equation}

All possibilities are accounted for by defining the {\it auxi\-lia\-ry 
map\/} $\Phi_{\e,a}$ to be the 2-parameter map (see figure 3)
\begin{equation} \label{Phi} 
\Phi_{\e,a}(x)=\left\{ \begin{array}{cll}
	f_+(x)&\quad\mbox{if }& x\in U_+\\[1.0ex]
	1-2\e &\quad\mbox{if }& x \in \Gamma\\[1.0ex]
	f_-(x)&\quad\mbox{if }& x \in U_-
\end{array}\right.\quad \ec < \e < 1-\ec. 
\end{equation}
It is plain that $\Phi_{\e,a}$ maps $U$ onto itself and if, in addition, 
$\gamma$ is positive, then $U_-$, $\Gamma$ and $U_+$ have positive 
measure. This characterizes the domain and range of $\Phi_{\e,a}$.

The auxiliary map reduces the dimensionality of our system. 
Originally the CML is an infinite dimensional system since 
it has an infinite number of sites. But in the case of minimal 
states the dynamics of the whole lattice is reduced to that
of the auxiliary map $\Phi_{\e,a}$, as illustrated schematically 
in figure 4. Iterating the CML $q$ times amounts to applying 
the auxiliary map $q$ times on the interface site. During these 
$q$ iterations, whenever the branch $f_-$ is used the interface 
is not shifted; whenever $f_+$ is applied we shift the 
interface by one site to the right; and finally, when the 
interface site falls into $\Gamma$ the CML essentially undergoes two 
iterations --- cf.~(\ref{TwoStages}) --- but the final result is 
that the interface site value is $1-2\e$ as well as being shifted
one site to the right.
\medskip

\subsection[4.2]{Symbolic dynamics}

The following binary symbolic dynamics for the 
auxiliary mapping $\Phi_{\e,a}$ will play a decisive role in the rest of 
the paper.

We assign the symbols `0' to $U_-$, the symbol `1' to $U_+$, and the 
symbol `10' to $\Gamma$. This choice originates from the prescriptions 
(\ref{ThreePossibilities}), according to which the location of a minimal 
state remains unchanged for $x\in U_-$, increases by one for $x\in U_+$, 
and increases by one for every two iterations for $x\in \Gamma$.
To the orbit of the auxiliary mapping with initial condition $x$ we 
associate the  semi-infinite sequence of symbols 
$$
S=S(x)=(s_1,s_2,s_3,\ldots\,) \qquad s_k\in\{0,1\}
$$
where we stipulate that the symbol `10' generates two binary
symbols: `10'$\longrightarrow$`1',`0'. For this reason, our code 
will not be unique. The space of binary sequences is denoted 
by $\Sigma_2$, and is equipped with the usual topology.

Any orbit with an element in $\Gamma$ will be called a {\it gap orbit}. 
If this orbit is periodic, then the period of its symbolic representation 
is one greater than that of the orbit, because the gap $\Gamma$ corresponds
to two symbols.

The {\it rotation number\/} \cite{Stark:88,Baker:book} of the orbit through 
$x$ is defined as 
\begin{equation} \label{RotationNumber}
\nu=\lim_{T\to\infty}{1\over T}\,\sum_{t=0}^T \,s_t
\end{equation}
provided that the limit exists.
If this is the case, it is plain that $0\leq \nu\leq 1$.

We will show that the symbol sequence $S(x)$ can be characterized 
in terms the {\it spectrum\/} of $\nu$, which is defined as the 
following sequence of integers
\begin{equation} \label{Spec}
Spec(\nu) \,=\, \lfloor \nu\rfloor,\lfloor 2\nu\rfloor,
		\lfloor 3\nu\rfloor,\,\dots 
	  \,=\, a_1,a_2,a_3,\,\ldots
\end{equation}
where $\lfloor\cdot\rfloor$ is the floor function. Because 
$0\leq \nu\leq 1$, consecutive elements of $Spec(\nu)$ differ 
by zero or one. Thus defining
$$
\tau(a_1,a_2,\dots\,)=(a_2-a_1,a_3-a_2,\dots\,),
$$
we find that $\tau(Spec(\nu))$ is a binary sequence. 

Using symbolic dynamics it is easy to follow the evolution of 
the minimal state: every iteration corresponds to a new symbol 
in the sequence. If the symbol is `0' then the minimal state remains 
in the same place and if the symbol is `1' the minimal state is 
shifted one site to the right. We can assume the velocity of the 
travelling front to be constant, so that the spectrum of the velocity
gives the position of the minimal state at every iteration, since
the location of the minimal state can only take integer values.
Thus $\tau$ applied on $Spec(\nu)$ gives a 0 when the position of 
the minimal state remains unchanged and a 1 when it jumps from on 
site to its right neighbour. Therefore the symbolic binary sequence 
associated to a given rotation number $\nu$ can be computed as
\begin{equation}\label{S(v)} 
 S=\tau\left(Spec(\nu)\right).
\end{equation}

If $\nu$ is rational, then $\tau(Spec(\nu))$ is periodic, as 
easily verified. For example, the sequence $S(3/7)$ is given by 
$$\begin{array}{rcl}
   S(3/7)&=&\tau\left(Spec(3/7)\right)\\[1.0ex]
	 &=&\tau(0,0,1,1,2,2,3,3,\,\ldots)\\[1.0ex]
	 &=&(\overline{0,1,0,1,0,1,0}),
\end{array}$$
where the bar denotes periodicity. 

The symbolic dynamics of $\Phi_{\e,a}$ is that of a uniform rotation 
with rotation number $\nu$ analogous to that of a subcritical circle 
map, as we shall see later.
The dynamics of the whole lattice is then reduced to the study of the
one-dimensional map $\Phi_{\e,a}$ whose rotation number gives the
velocity of propagation of the travelling front. Using arguments similar
to that given in section 1.14 of \cite{Devaney:book},
it is possible to prove
that the rotation number is a well-defined, non-decreasing, continuous
function of the parameters $(\e,a)$ and it is independent of the
initial condition. The difference here is that the auxiliary map is a
non-decreasing function and not a strictly increasing one as in
\cite{Devaney:book}, but the same line of proof may be followed. Therefore,
when the gap size is non-negative, the velocity $v(\e)$ of the travelling
interface is a well defined, non-decreasing, continuous function and
does not depend on the choice of the initial minimal state.\par
\bigskip

\section[5]{Structure of parameter space}

We now turn to the study of the parameter space of the coupled map 
lattice $F_{\e,a}$, in the case where the gap size $\gamma(\e,a)$ 
is non-negative (cf.~(\ref{GapSize})). Then the gap orbit is defined, 
and we denote by $v=v(\e,a)$ the velocity of the center of mass of the 
corresponding minimal state.
\medskip

\subsection[5.1]{Mode-Locking}

Using the symbolic coding we now prove that the gap $\Gamma$ 
forces the velocity of the travelling wave to be locked to 
rational values if $\Phi_{\e,a}$ has an eventually periodic orbit 
containing the point $1-2\e$. Mode-locking guarantees stability of the 
propagating signal with respect to parametric perturbations.

The symbolic coding of the orbit of the auxiliary map $\Phi_{\e,a}$ 
gives the velocity of the travelling interface via the rotation number 
(\ref{RotationNumber}). The presence of a non-negative gap $\Gamma$ 
induces a mode-locking of the rotation number in the following way. Let us 
consider the orbit generated from an arbitrary point $x_0$ in $\Gamma$. If 
after $q$ iterations $x_q$ falls into $\Gamma$, the orbit is eventually 
periodic \cite{Devaney:book} of period $q$ with rotation number $p/q$, $p-1$ 
being the number of times the orbit visits $U_+$ (recall that there is an extra 
symbol `1' coming from the gap). 

Suppose now that we introduce a small perturbation of the parameters and 
the initial condition: $(x_0,\e,a)\longrightarrow(x_0',\e',a')$. The 
perturbed orbit $x'_t=\Phi^t_{\e',a'}(x_0')$ is initially at a distance 
$\Delta x_0=Ýx_0'-x_0Ý$ from the unperturbed one. Using the continuity 
of $x'_t$ on the parameters $\e,a$ and the initial condition, the distance 
$\Delta x_t$ between the two orbits at time $t$ can be made as small
as we want by making $(x_0,\e,a)$ sufficiently close to $(x_0',\e',a')$.
Because $\Gamma$ is an open interval, if $x_q$ belongs to $\Gamma$, then 
so does $x'_q$ for a sufficiently small perturbation. Therefore we have 
established the crucial result:
\medskip

\proclaim Theorem {3}. If the gap orbit of $\Phi_{\e,a}$ is finite, then 
it is stable under a sufficiently small perturbation of parameters and 
initial condition.
\medskip

Thus there is a region in the parameter space $(\e,a)$, where the given 
rotation number is constant or {\it mode-locked}. An example of mode-locking 
$\e$-regions for a fixed value of $a$ is given in figure 1. The mode-locked 
region, corresponding to a given rotation number $\nu=p/q$, can be computed 
by noting that in order that an orbit through any point in $\Gamma$ returns 
after $q-1$ iterations we must have that 
$\Phi_{\e,a}^{q-1}(\Gamma)\subseteq\Gamma$. This condition may be rewritten 
as $\Phi_{\e,a}^{q-2}(1-2\e) \in \Gamma$ giving the inequalities
\begin{equation} \label{PhiGammaInGamma}
\gamma_+\quad \leq\quad\Phi_{\e,a}^{q-2}(1-2\e)\quad\leq\quad\gamma_-,
\end{equation}
since 
$$\forall\, x_0\in\Gamma\quad\Longrightarrow\quad x_1=\Phi_{\e,a}(x_0)=1-2\e.$$
We shall mention that the end points of (\ref{PhiGammaInGamma}) 
are included because an orbit arriving at $\gamma_\pm$ after 
$q-2$ iterations will also reach $1-2\e$ after two more iterations 
since $\Phi_{\e,a}\left(f_\pm(\gamma_\pm)\right)=\Phi_{\e,a}(\pm a)=1-2\e$ 
(cf.~(\ref{LotsOfValues})) --- these two orbits correspond to the
kneading sequences of the local map \cite{Sparrow:90,Sparrow:93}. 
The closed $\e$-interval region for which the velocity is mode-locked 
to a given rational is called a {\it plateau}.

Equation (\ref{PhiGammaInGamma}) ensures that the orbit starting 
at $x_1=1-2\e$ falls into the gap after $q-2$ iterations and 
therefore repeats itself after $q-1$ iterations. The value of 
$\Phi_{\e,a}^{q-2}(1-2\e)$ is determined by the coding sequence of the 
associated rotation number. So if we want to compute the mode-locked 
region for a given velocity $\nu=p/q$ we have to determine the 
$(\e,a)$ region where (\ref{PhiGammaInGamma}) is satisfied in 
correspondence to the code $S(p/q)$ given by (\ref{S(v)}). For 
example, to compute the mode-locked region for the velocity $\nu=2/5$ 
we first compute $S(2/5)=(\overline{0,1,0,1,0})$, so we know that 
the orbit visits the regions $U_-,U_+,U_-$ and $\Gamma$. Solving for 
$\Phi_{\e,a}^3(1-2\e)=f_-\left(f_+\left(f_-(1-2\e)\right)\right)$ gives
$$
\gamma_+\quad \leq\quad f_-\left(f_+\left(f_-(1-2\e)\right)\right)
\quad \leq\quad \gamma_-.
$$
In figure 5 we show the mode-locking regions in the $(\e,a)$-plane,
computed via (\ref{PhiGammaInGamma}), for some rational velocities.
These regions, represented as shaded areas, are the so-called
{\it Arnold's tongues} \cite{Baker:book,Bak:86}. It can be observed 
that as we approach the zero-gap line (dashed line) --- given by
$\gamma=0$ in (\ref{GapSize}) --- the tongue size decreases, as the 
latter is related to the gap size.
\medskip

\subsection[5.2]{Lattice representation}

We now introduce a geometric representation of the rational velocities 
of the wavefront. This representation gives a straightforward method 
for computing the associated coding sequence as well as some properties
related to the mediant of two given consecutive rational velocities of a 
Farey sequence.

We represent the rational velocity $v=m/n$ as the point $P(m/n)=(n-m,m)$ 
on the integral lattice ${\Bbb Z}^2$. Consider the ray $O\!P$ emanating 
from the origin and passing through $P(m/n)$. It is obvious that distinct 
irreducible rational velocities correspond to distinct points and distinct 
rays. The ray $O\!P$ will intersect the lattice grid infinitely many times. 
We drop the first intersection in $O$ and take all the remaining ones. 
Next, write a `0' every time the ray $O\!P$ crosses a vertical grid line, 
write a `1' when it crosses a horizontal grid line and write `10' when its 
crosses through a lattice point (see example in figure 6). The resulting 
semi-infinite binary sequence can be shown to be equal to the symbolic 
coding of the velocity $\nu=m/n$ given by (\ref{S(v)}). 

We sketch the proof. Recall that a `0' in $S(m/n)$ means one 
iteration of the CML without advance of the interface while a `1' means 
that the interface has advanced one site. Let us now consider the 
consequence of crossing vertically and horizontally the grid lines of 
${\Bbb Z}^2$. Crossing a vertical line of the grid means transition from 
the point $(n-m,m)$ to the point $(n+1-m,m)$, {\it i.e.}\ from $P(m/n)$ 
to $P(m/(n+1))$ --- this corresponds to iterating the CML once without any 
advance of the interface (symbol `0'). Crossing a horizontal line means 
moving from the point $(n-m,m)$ to the point $(n-m,m+1)=((n+1)-(m+1),m+1)$,
{\it i.e.}\ from $P(m/n)$ to $P((m+1)/(n+1))$, which corresponds to 
a shift of one site of the interface after one iteration (symbol `1'). And 
finally, crossing through a lattice point is equivalent to a shift of one 
place of the interface in two iterations (symbol `10').

We recall that the Farey series \cite{Knuth:book,Hardy:book} of order $r$, 
{\eusm F}$\!_{r}$, is defined 
to be the set of irreducible fractions, in ascending order, belonging to 
$[0,1]$ whose denominators are smaller than or equal to $r$. Suppose now 
that the velocities $v=m/n$ and $v'=m'/n'$ are two consecutive rationals 
of the Farey series {\eusm F}$\!_{\max(n,n')}$. Then, using the 
lattice representation, it is not difficult to see that if $m/n < m'/n'$ 
the coding sequence of the velocity $v''=(m+m')/(n+n')$ corresponding to 
their {\it mediant\/} is given by
\begin{equation} \label{AddSequences}
S\left(m+m'\over n+n'\right)=
S\left({m\phantom{'}\over n}\!\right) S\left(m'\over n'\right),
\end{equation}
where the right hand side denotes the {\it concatenation\/} of two 
periodic sequences, defined as follows. If $S_a=(\overline{a_1,\dots,a_p})$ 
and $S_b=(\overline{b_1,\dots,b_q})$ are periodic sequences of period $p$ and 
$q$, respectively, their concatenation $S_a S_b$ is the periodic sequence of 
period $p+q$ defined as
\begin{equation} \label{Concatenation}
S_a S_b=(\overline{a_1,\dots,a_p,b_1,\dots,b_q}).
\end{equation}
The result (\ref{AddSequences}) derives from the fact that the 
parallelogram $(O\!P\!P''\!P')$, with $P=P(m/n), P'=P'(m'/n')$ 
and $P''=P''(m''/n'')=P''((m+m')/(n+n'))$, does not contain any 
lattice point since $m/n$ and $m'/n'$ are consecutive numbers of 
a Farey series \cite{Hardy:book}.

An example of this construction is shown in figure 6 where the velocities 
$v=1/3$ and $v'=1/2$ give the mediant $v''=2/5$. The sequence for $v=1/3$ 
is ($\overline{010}$) (vertical cross + lattice point) and the sequence 
for $v'=1/2$ is ($\overline{10}$) (just a lattice point) and the resulting 
sequence for the mediant ($\overline{01010}$) is obtained by the 
concatenation $(\overline{010})(\overline{10})$.
\medskip

\subsection[5.3]{Unimodular transformations and en\-ve\-lo\-pes}
The structure of the parameter space of the mapping $F_{\e,a}$ at 
the boundary of a tongue can be described analytically by means of 
{\it envelopes}. These are functions characterizing the structure of 
sequences of adjacent tongues. The study of envelopes proceeds in two 
stages. Firstly we derive upper and lower envelopes for the so-called 
first order plateaus ({\it i.e.}\ $v=1/n$ and $v=(n-1)/n,\ n=1,2,\dots$). 
Then with the use of unimodular transformations we find envelopes for 
any family of plateaus.

The {\it zeroth order plateaus\/} are defined as the two plateaus given by 
(\ref{ZerothPlateaus}) with velocities \{$0/1,1/1$\}. In order to construct 
any order of plateaus, the {\it $k$-th order plateaus}, we take 
any two consecutive elements of the previous order $k-1$ denoted by 
$v_a=m_a/n_a$ and $v_b=m_b/n_b$ ($v_a<v_b$). Then the next order of 
plateaus is defined as the following infinite increasing sequence of mediants
\begin{equation}\label{NextOrder}
\left\{\cdots,
{\ds 3 m_a+m_b\over\ds 3 n_a+n_b},
{\ds 2 m_a+m_b\over\ds 2 n_a+n_b},
{\ds m_a+m_b\over\ds n_a+n_b},
{\ds m_a+2 m_b\over\ds n_a+2 n_b},
\cdots\right\}
\end{equation}
The first order plateaus are thus given by
$$
\left\{\cdots,
{\ds 1\over\ds 4},
{\ds 1\over\ds 3},
{\ds 1\over\ds 2},
{\ds 2\over\ds 3},
{\ds 3\over\ds 4},
\cdots\right\}.
$$
They form an unique family. On the other hand, between any two consecutive 
plateaus of a given order there is a new family of plateaus of the next 
order and so on. This construction is similar to that of the Farey series
\cite{Knuth:book}.
In particular, any two consecutive plateaus $p/q$ and $p'/q'$ ($p/q<p'/q'$) 
of the same order and family are consecutive fractions of a Farey series
({\it i.e.}\ $qp'-pq'=1$). 

From the definition of the sequence of plateaus (\ref{NextOrder}) and by 
applying repetitively the sequence concatenation (\ref{AddSequences}) it 
is possible to find the sequences associated to a given family of plateaus:
$$
\left\{\dots,\,S_a^3 S_b,\,S_a^2 S_b,\,S_a S_b,\,S_a S_b^2,\,
S_a S_b^3,\dots\right\},
$$
where we used the shorthand notation $S_a=S(m_a/n_a)$ and $S_b=S(m_b/n_b)$. 
It will be useful to distinguish the left $\Sigma_a$ and right $\Sigma_b$ 
subfamilies of sequences defined by
\begin{equation}\label{Sigma_ab} \begin{array}{rcl}
\Sigma_a&=&\left\{\dots,\,S_a^3 S_b,\,S_a^2 S_b,\,S_a S_b\right\}\\[1.0ex]
\Sigma_b&=&\left\{S_a S_b,\,S_a S_b^2,\,S_a S_b^3,\dots\right\}.
\end{array}\end{equation}
Note that the sequence $S_a S_b$, corresponding to the mediant of $v_a$ 
and $v_b$, appears in both families. The elements of the subfamilies
$\Sigma_a$ ($\Sigma_b$) tend, to the left (right), to $S_a$ ($S_b$). 
Thus the subfamilies $\Sigma_a$ ($\Sigma_b$) correspond to velocities 
going from the mediant of $v_a$ and $v_b$ to the plateau $v_a$ ($v_b$).

From now on we consider only $\Sigma_a$. All the results generalize to
$\Sigma_b$ just by interchanging the subscripts $a$ and $b$. The elements 
of $\Sigma_a$ are of the form
\begin{equation}\label{Sigma_a} 
S\in\Sigma_a \quad\Longrightarrow\quad S=S_a^n S_b\qquad n=1,2,\dots
\end{equation}
which correspond to the velocities
\begin{equation}\label{v(Sigma_a)} 
v(S_a^n S_b)={\ds n\, m_a+m_b \over n\, n_a + n_b} \qquad n=1,2,\dots
\end{equation}
We recall that the transformations (\ref{v(Sigma_a)}) are unimodular 
since we always deal with consecutive fractions of Farey series.

The sequences $S_a$ and $S_b$ correspond to a particular combination
of $f_-$ and $f_+$. Since $f_\pm$ are linear, any such combination 
will again give a linear function. We denote by $f_{S_a}$ ($f_{S_b}$) 
the linear function resulting from the combination of $f_\pm$ specified
by the sequence $S_a$ ($S_b$):
\begin{equation} \label{f_S} \begin{array}{lcr}
f_{S_a}(x) &=& \alpha_a x +\beta_a \\[1.0ex]
f_{S_b}(x) &=& \alpha_b x +\beta_b.
\end{array} \end{equation}
Composing $f_{S_a}$ with itself $n$ times gives
\begin{equation} \label{f^n(S_a)}
f^n_{S_a}(x) = f_{S^n_a}(x) =\alpha_a^n(x-\xi_a^*) + \xi_a^*
\end{equation}
where $\xi_a^*\equiv \beta_a/(1-\alpha_a)$ is the fixed point of $f_{S_a}$.
Combining (\ref{f_S}) and (\ref{f^n(S_a)}) in (\ref{Sigma_a}) we obtain the 
iterates of $x_0=1-2\e$ under $\Phi_{\e,a}$ for the coding sequences $S_a^nS_b$
\begin{equation} \label{f(Sigma_a)}
f_{S_a^nS_b}(x_0) = f_{S_b}\left(f^n_{S_a}(x_0)\right).
\end{equation}
Replacing (\ref{f(Sigma_a)}) in (\ref{PhiGammaInGamma}) we obtain the
mode-locking region for a given series of plateaus
\begin{equation} \label{Plateaus}
f_{10}(\gamma_+)\quad \leq \quad
\alpha_b (\alpha_a^n(x_0-\xi_a^*)+\xi_a^*)+\beta_b
\quad \leq \quad f_{10}(\gamma_-),
\end{equation}
where we have taken into account the fact that the two final symbols `10' 
of any sequence in $\Sigma_a$ come from the gap $\Gamma$ and therefore they 
have to be compensated by replacing $\gamma_\pm$ in (\ref{PhiGammaInGamma})
by $f_1(f_0(\gamma_\pm))=f_{10}(\gamma_\pm)$ .

As we mentioned earlier, the plateaus include their end points. These are 
obtained from solving the inequalities (\ref{Plateaus}) for the extremes 
$f_{10}(\gamma_\pm)$. Therefore, all the end points of a given family of 
plateaus are given by
\begin{equation} \label{EndPoints}
\alpha_b (\alpha_a^n(x_0-\xi_a^*)+\xi_a^*)+\beta_b=f_{10}(\gamma_\pm),
\end{equation}
where the positive (negative) sign gives the left (right) end 
points of the plateaus. As we vary $n=0,1,\dots$ in (\ref{EndPoints}) 
we browse all the extremes of the series of plateaus going from the 
mediant of $v_a$ and $v_b$ to $v_a$.

Solving (\ref{EndPoints}) for $n$ yields
$$
n_\pm(\e,a)={\ds 1\over \ds \ln\alpha_a}
\,\ln {\ds f_{10}(\gamma_\pm) -\beta_b-\alpha_b \xi_a^* \over 
\ds \alpha_b(x_0-\xi_a^*)}.
$$
Finally, replacing the value of $n$ in (\ref{v(Sigma_a)}) by $n_\pm(\e,a)$ 
gives the continuous functions
\begin{equation} \label{Envol} 
v_\pm(\e,a)={\ds n_\pm(\e,a)\, m_a+ m_b \over 
\ds n_\pm(\e,a)\, n_a + n_b},
\end{equation}
that passes through all the left ($v_+$) and right ($v_-$) boundaries of 
the plateaus. Therefore, $v_\pm$ gives the upper ($+$) and lower ($-$)
envelopes of the series of plateaus (\ref{v(Sigma_a)}).

In figure 7a we display the first order plateaus for $a=0.4$. We only 
plotted the left family ($v\leq 1/2$) since the right family ($v\geq 1/2$) 
is symmetrical from (\ref{Symmetry}). An example of a second order family, 
for the same value of $a$, is shown in figure 7b where we display the 
lower and upper envelopes for the left and right families of second 
order plateaus between $v=1/2$ and $v=1/3$.

With the method described above, it is possible to find the envelopes 
of any sequence of high order plateaus via unimodular transformations. 
The self-similarity structure of the velocity function, easily observable 
in figures 7a and 7b, is controlled by modular transformations 
(\ref{v(Sigma_a)}) and envelopes. Since it is possible to find a periodic 
symbolic sequence for any given rational velocity arising from an eventually 
periodic orbit of $\Phi_{\e,a}$ through $1-2\e$ then it is possible to 
find a non-empty $\e$ interval where the velocity is mode-locked (Theorem 3).
Therefore the graph of the velocity $v(\e,a)$ as a function of $\e$ is in 
fact a {\it Devil's staircase\/} \cite{Bak:86,Schroe:book}, 
{\it i.e.}\ a fractal staircase. 
This mode-locking phenomenon is very similar to the one observed for a 
uniform rotation in a subcritical perturbed circle map \cite{Baker:book}.

If, on the other hand, the symbolic sequence is chosen to be non periodic,
the associated velocity $\nu$ is irrational. The continuity of the function
$v(\e)$ guarantees that there exists at least one value of $\e$ such that
$v(\e)=\nu$. We believe that it is possible to prove that there is only one
such $\e$.  The idea is to take a sequence of rational approximants of
$\nu$; the corresponding upper and lower envelopes should approach
each other, as the approximants tend to $\nu$, and coincide in the limit to
give a unique $\e$ such that $v(\e)=\nu$.
\medskip

\subsection[5.4]{Convergence to minimal states}

In this chapter we have described the structure of the parameter space
associated with minimal states. Here we generalize the notion of minimal
state and briefly complete our description of the parameter space.

A minimal state --- a step state of size 0 or 1 --- is stable under 
iteration when the gap is non-negative. When the gap is negative, the 
stable configuration of a step state is found to have size assuming only
two values, $N$ and $N-1$. Such states will be called minimal $N$-states.

Numerical experimentation reveals that the integer $N$ does not depend
on the initial condition. Any initial configuration we have tried in 
the interface (increasing, decreasing, oscillatory, random with small 
and large size, etc \dots) converged, after a transient (depending on 
the initial conditions) to a minimal $N$-state, with the value of $N$
depending only on $\e$ and $a$. Accordingly, the parameter space is 
partitioned into layers where the value of $N$ is constant, as depicted
in figure 8. The boundary between the first layer (minimal 1-state) 
and the second one (minimal 2-state) is the zero-gap curve, given by
$\gamma=0$ in (\ref{GapSize}). Above that curve the gap is negative, and 
there appears to be an infinite family of layers labeled by the value 
of $N\!$, separated by boundary curves which may not be differentiable.

The dynamical analysis for such minimal $N$-states may be carried out by
reducing the dynamics of the whole lattice to a $N$-dimensional auxiliary map
\cite{rcg:97} analogous to the one obtained previously for minimal 1-states.
\bigskip

\section{Generalizations and conclusions}

We have described a mechanism that allows travelling interfaces in a 
one-way CML (\ref{one_way}) for a special family of piece-wise linear 
maps (\ref{fa}). We now give some ideas on the generalization of these 
results to a broader range of functions and couplings.

First of all, the superattractiveness of the points $x_-^*$ and $x_+^*$ 
is not necessary. It suffices that the map $f$ be a contraction mapping at
$x_\pm^*$. This is the case, for instance, if $f''(x)<0$ for $x_-^*<x<x^*$ 
and $f''(x)>0$ for $x^*<x<x_+^*$, where $x^*$ is the unstable fixed point 
in between $x_-^*$ and $x_+^*$. However, without superattractiveness, the 
travelling wave will be less stable to perturbations. Then, if we are dealing 
with a medium involving random fluctuations we would like the map to be 
sufficiently contracting for a travelling interface to survive.
On the other hand, the lack of superstability of $x_\pm^*$ introduces more
sites in the interface since the convergence to $x_\pm^*$ is slower. Moreover,
in this case the
number of sites contained in the open interval $(x_-^*,x_+^*)$ is not
necessarily finite (take the example of an exponentially localized state).
For this reason the reduction of the dynamics in this case does not lead to
any simplification. Nevertheless, if the shape of the front remains unchanged
during the evolution, it is possible to concentrate our attention to a single
site in the interface since the remaining sites are determined by the shape
of the front. We could, for example, consider the site
that is closest to the unstable point $x^*$ and from it reconstruct a
one-dimensional auxiliary map that accounts for the whole dynamics, in the
same way that $\Phi_{\e,a}$ accounts for the whole dynamics of minimal
states. A more detailed study will be given in \cite{rcg:97}.

Secondly, the discontinuity of the derivative of the local map is not 
necessary for mode-locking. We have observed mode-locking for analytic 
maps as well. In figure 9 a window of the velocity as a function of $\e$ 
is shown for some polynomials maps (of odd degrees from 5 to 13). The 
plateaus $v=1/2$ and $v=1/3$ are clearly visible. The origin of mode-locking
is not related to non-analyticity. It ultimately lies with the asymptotic 
($t\rightarrow\infty$) discontinuity from the initial conditions associated 
with the repelling fixed point $x^*$ between the two stable points, which 
separates two dynamical behaviours.

In this respect the value of the derivative at the repeller can be
interpreted as a measure of this discontinuity. It is therefore 
expected that, for fixed value of the derivative at the attractors, 
the mode-locked plateaus become more pronounced as the derivative 
at the repeller point increases. In figure 9 the plateaus $v=1/2$ 
and $v=1/3$ are almost imperceptible for the polynomial of degree 5 
that has a derivative at the repelling point $\sim\!\!1.9$, meanwhile 
they become more appreciable for the polynomials of higher degree 
(7, 9, 11 and 13) where the derivative at the repelling point gets 
larger ($\sim\!\!2.2$, 2.5$, 2.7$ and $2.9$ 
respectively). 

Finally, the travelling wave property arises from one-way coupling 
in the  CML and the direction of the wave is given by that of the 
coupling. By comparison, a front on a diffusive CML (\ref{diffu}) 
with a symmetric map at each site is going to remain stationary in 
the long term. For the front to advance one needs asymmetry, either 
on the coupling or in the local map (with respect to the repeller 
point $x^*$) which generates a bias between competing attractors.

In order to illustrate this phenomenon let us take the logistic map 
$f_\mu (x)=\mu x(1-x)$. The logistic map itself does not fulfill the 
requirements (two attractive points with a repeller in between) but 
consider its second iterate $f^2_\mu$. At the parameter value 
$\mu=1+\sqrt{5}$ the mapping $f^2$ has two superattractive fixed points 
$x_-^*=1/2$ and $x_+^*=1/2+\sqrt{2}\sqrt{3-\sqrt{5}}/4)$ (corresponding 
to a superattractive 2-cycle for $f$) separated by a repeller 
$x^*=1-\sqrt{2}\sqrt{3-\sqrt{5}}/4$. Consider iterating the diffusive 
CML with the map $f^2_\mu$ at each site. For an initial step state 
we  obtain a travelling interface where the velocity again behaves 
as in the examples shown above --- it has a transition from $v(\e)=0$ 
to $v(\e)>0$ at $\ec\simeq0.198$ ---, see figure 10. In this case the 
mode-locking is not apparent at first glance. However, on amplification, 
the plateaus are shown to occur. For example, the plateaus corresponding 
to $v=1/7$ (right) and $v=1/8$ (left) are shown in the enlargements in 
figure 10.

In conclusion, the mode-locking of the velocity of the travelling 
interface, appears to be an universal phenomenon in coupled map 
lattices. It implies that waves travelling with rational velocities 
are stable under both spatial and parametric perturbations.
\bigskip

\appendix
\def\theequation{A.\arabic{equation}}
\setcounter{equation}{0}
\noindent{\Large\bf Appendix A}
\medskip

In this appendix we provide various statements and proofs concerning
the image of localized states.

\proclaim Lemma {A.1}. %
Let $X_t$ be a localized state of a one-way CML, let 
$x_\pm=\lim_{i\to\pm\infty} x_t(i),$ and let $M(X_{t+1})\neq 0$.
Then, if $f$ is bounded and continuous at $x_\pm$, $X_{t+1}$ is 
a localized state. If, in addition, $X_t$ is exponentially localized, 
so is $X_{t+1}$.

\noindent {\it Proof.} We begin considering the case $i\to\infty$. Because 
$X_t$ is localized, $x_+$ is finite. Moreover, since $f$ is continuous at 
$x_+$, then there exists a constant $\rho>0$ and an integer $N$ such that, 
for all $i\geq N$ we have $|f(x_t(i+1))-f(x_t(i))|<\rho |x_t(i+1)-x_t(i)|$,
whence
\begin{equation} \label{Inequality}
\begin{array}{rcl}
\Delta x_{t+1}(i) 
\,&=&\,   |(1-\varepsilon)\,f(x_t(i+1))+\varepsilon \,f(x_t(i))\\[1.0ex]
 & &     -(1-\varepsilon)\,f(x_t(i))-\varepsilon \,f(x_t(i-1))| \\[1.0ex]
\,&\leq&\,(1-\varepsilon)\,|f(x_t(i+1))-f(x_t(i))|\\[1.0ex]
 & &     +\, \varepsilon \,|f(x_t(i))-f(x_t(i-1))|\\[1.0ex]
\,&<&\,\rho\,(1-\varepsilon)\,|x_t(i+1)-x_t(i)|\\[1.0ex]
 & &     +\, \rho \varepsilon \,|x_t(i)-x_t(i-1)|\\[1.0ex]
\,&=&\,\rho\,(1-\varepsilon)\,\Delta x_{t}(i)\,+
\,\rho\varepsilon\Delta x_{t}(i-1), \quad i>N. 
\end{array}
\end{equation}
Then, from absolute convergence
$$
\begin{array}{c}
\sum_{i=N+1}^\infty \Delta x_{t+1}(i)
 \,<\, \rho (1-\varepsilon)\,\sum_{i=N+1}^\infty \Delta x_{t}(i) \\[1.0ex]
\,+\, \rho \varepsilon\,\sum_{i=N+1}^\infty \Delta x_{t}(i-1) < \infty.
\end{array}
$$
A similar inequality holds for $i\to-\infty$. From the above and the 
fact that any finite sum of $\Delta x_{t+1}(i)$ is bounded (because
$f$ is bounded) it follows that $M_{t+1}$ is finite (and non-zero, 
by hypothesis), so that the probability $p_{t+1}$ exists. Finally, 
multiplying both sides of the inequality (\ref{Inequality}) by $i$ and 
by $i^2$, respectively, and summing over $i>N$ shows that $\mu_{t+1}$ 
and $\sigma_{t+1}$ are finite, whence $X_{t+1}$ is localized. 

If $X_t$ is exponentially localized, then there exist positive constants 
$c$ and $\kappa<1$ for which
\begin{equation}\label{Bound}
\Delta x_t(i)<c\kappa^i.
\end{equation}
Combining the above with (\ref{Inequality}) we obtain
\begin{equation}\label{PositiveI}
\Delta x_{t+1}(i) \,<\,
  \kappa^{i}c\rho\left((1-\varepsilon)+{\varepsilon\over\kappa}\right),
  \qquad\quad i>N.
\end{equation}
A similar estimate holds for negative $i$, showing that $X_{t+1}$ is 
exponentially localized (possibly with a larger constant $c$).\hfill$\Box$
\bigskip

From Lemma A.1 and the fact that the mass of a step state is bounded 
from (cf.~(\ref{MinimalMass})) we obtain immediately

\proclaim Corollary {A.2}. %
Let $X_t$ be a step state of a one-way CML, and let $f$ be continuous at 
$x^*_\pm$ and bounded. Then $X_{t+1}$ is a step state. If, in addition, 
$X_t$ is exponentially localized, so is $X_{t+1}$.
\bigskip

We have then establish the invariance of the localized and exponentially
localized states under the dynamics of the CML and we are ready now to 
give a proof for Theorem 1.

\proclaim Theorem 1. %
Let $X_0$ be an exponentially localized step 
state of a one-way CML. Let the local map $f$ be bounded, and let 
$f$ be a contraction mapping in a neighbourhood of the fixed points 
{\rm (\ref{FixedPoints})}. Then, for all sufficiently small $\e$,
$v(\e)=0$ and $v(1-\e)=1$, independently of $X_0$.

\noindent {\sl Proof.} 
We deal with the case of small coupling first, and positive $i$. From 
Corollary A.2 we know that $X_t$ is an exponentially localized step 
state for all $t\geq 0$. By assumption $f$ is a contraction mapping in 
some domain $|x-x^*_+|<r$. Let $N'$ be such that for all $i\geq N'$ we 
have $|x_t(i)-x^*_+|<r/2$, so that $\Delta x_t(i)<r$. Then, for $i\geq N'$, 
the constant $\rho$ appearing in (\ref{Inequality}) can be chosen to be 
smaller than 1.

Let $N''$ be such that $\Delta x_t(i)<1$ for all $i\geq N''$, and let 
$N={\rm max}(N',N'')$. Then since $X_t$ is exponentially localized, we 
can choose $\kappa$ so that the bound (\ref{Bound}) holds for $i\geq N$ 
with $c=1$. Letting $c=1$ in (\ref{PositiveI}) it follows that
\begin{equation}\label{TplusOneBound}
\Delta x_{t+1}(i)\,<\,\kappa^i\qquad\quad i> N
\end{equation}
provided that
$$
\varepsilon < {\kappa \over \rho}\,{1-\rho\over 1-\kappa}
$$
which is satisfied for sufficiently small $\varepsilon$, since the right 
hand side is positive.

It remains to consider the case $i=N$. Because $f$ is bounded, the quantity
$$
\Delta f = {\rm sup}_xf(x) - {\rm inf}_{x}f(x)
$$
is finite. We have, in place of (\ref{Inequality})
$$
\Delta x_{t+1}(N) < \rho\,(1-\varepsilon)\,\Delta x_t(N)+\varepsilon \,\Delta f
$$
giving, for all $t\geq 0$
$$
\Delta x_{t+1}(N) < \alpha^{t+1}\kappa^N+\varepsilon\,\Delta f\,
{1-\alpha^{t+1}\over 1-\alpha},
$$
where $\alpha=\rho(1-\varepsilon)<1$. The right-hand side of the 
above inequality can be made smaller than $\kappa^N$ for all $t$, 
provided that $\varepsilon$ is sufficiently small. Thus the inequality 
(\ref{TplusOneBound}) holds also for $i=N$.

Let $i$ be negative. Assuming that $\Delta x_t(i)<\kappa^{-i}$ for all 
sufficiently large (negative) $i$, and proceeding as above, we find 
that the bound $\Delta x_t(i)<\kappa^{-i}$ for $i\leq -N$ implies that
$$
\Delta x_{t+1}(N) < \kappa^{-i}\rho\,(1-\varepsilon+\varepsilon\kappa)
  < \kappa^{-i}; \qquad\quad i < -N
$$
for all $\varepsilon$ (all constants have been redefined). The case $i=-N$ 
gives the recursive inequality
$$
\Delta x_{t+1}(-N) < (1-\varepsilon)\,\Delta f+\rho\varepsilon \kappa^{-N+1}
$$
which, as before, yields $\Delta x_{t+1}(-N)<\kappa^{-N}$ for sufficiently
small $\varepsilon$.

The above induction establishes the time-independent bound
$$
\Delta x_{t}(i) < \kappa^{|i|}, \qquad\qquad {|i|}\geq N, \qquad t\geq 0.
$$
for a suitable choice of $\kappa$ and $N$, and for all sufficiently
small $\varepsilon$. In this $\varepsilon$-range, from the boundedness of 
$f$ we conclude that $|\mu_t|$ and $\sigma_t$ are bounded from above 
for all times. Thus $v(\varepsilon)=0$, and $X_t$ remains a step state 
also in the limit $t\to\infty$. $\hfill\Box$
\bigskip

\appendix
\def\theequation{B.\arabic{equation}}
\noindent{\Large\bf Appendix B}
\medskip

\noindent {\sl Proof of Theorem 2.} 

$(i)$ The dynamics of the $i$-th site is given by $x_{t+k}(i)=f_-^k(x_t(i))$ 
since it is coupled to the $(i-1)$-th site whose value is $-1$. For the 
dynamics of the $(i+1)$-th site, since $\e \leq \ec$ and $a\leq f_+(x)$ 
for any $x\in U$, we have that $x_{t+k}(i+1)\in S_+$. Therefore at time 
$t+k$ the state of the lattice will be equivalent to $X_{t+k}=[f_-^k(x_t),i]$ 
which depends exclusively on the initial condition $x_t(i)$. If $\e \leq \ec$ 
the derivative of the map $f'_-(x)=(1-\e)/a\geq 1$ and then $f_-$ has a 
repeller at $x_-$. Therefore $f_-^k(x_t(i))$ will decrease (increase), 
for $x_t(i)<x_-$ ($x_t(i)>x_-$) until it reaches $S_-$ ($S_+$) when the 
state of the lattice is $P(i)$ ($P(i-1)$). For the marginal case $x_t(i)=x_-$ 
the state of the lattice $X_{t+k} = [f_-^k(x_-),i]=[f_-(x_-),i]$ is fixed 
and unstable since $x_-$ is an unstable fixed point of $f_-$.

$(ii)$ Using a similar reasoning, if $\e\geq 1-\ec$, $x_+$ is an 
unstable fixed point of $f_+$ and the state of the lattice at time 
$t+k$ is $X_{t+k}=[f_+^k(x_t),i+k]$. Therefore the lattice will 
reach the state $P(i+k)$ ($P(i+k-1)$) for $x_t(i)<x_+$ ($x_t(i)>x_+$). 
In the marginal case $x_t(i)=x_+$, the state of the lattice is 
$X_{t+k}=G^k\left([f_+^k(x_+),i+k]\right)= G^k\left([f_+(x_+),i]\right)$, 
which is fixed and unstable under $G\circ F$.\hfill $\Box$
\bigskip

\noindent{\Large\bf Acknowledgments}

RCG would like to thanks DGAPA for financial support during the 
preparation of this paper and Carmen Cisneros and Alvaro Salas Brito
for all their inconditional support.

\end{document}